# Have Large Language Models "Enhanced" the Way Civil & Environmental Engineers Write?
# A Quantitative Analysis of Scholarly Communication over 25 Years

Morgan D. Sanger[1] and Brett W. Maurer, Ph.D.[2]

**Abstract**

Large language models (LLMs) have rapidly emerged in civil and environmental engineering (CEE) research, education, and practice as tools for project ideation, execution, and communication. However, a gap remains in understanding how prevalent LLM adoption is across CEE scholarship and whether it measurably alters research prose. This study applies a vocabulary-based frequency-shift methodology to 149,452 abstracts published by the American Society of Civil Engineers between 2000 and 2025. Deviations from long-term vocabulary trends suggest that 15.3% and 26.9% of 2024 and 2025 abstracts, respectively, are likely LLM-influenced. Prior to the introduction of LLMs in 2022, CEE publications exhibit long-term trends toward longer abstracts, greater use of segmenting punctuation, higher required reading levels, and use of active, first-person verb constructions. Beginning around 2023, however, the frequencies of many stylistic marker words (e.g., enhance) sharply depart from their historical trajectories, a shift consistent with the emergence of LLMs. Abstracts classified as likely LLM-influenced also depart from historical semantic trends, exhibiting increased lexical diversity, comma use, and complexity, with reduced passive voice and hedging language, producing prose that is more segmented, complex, and confident. The artificial intelligence contribution of this study is the use of natural language processing to detect population-level linguistic signals of LLM-generated text, estimating the prevalence of LLM use and its influence on CEE writing. Together, these novel findings provide the first large-scale, data-driven assessment of how CEE scholarly communication is changing alongside LLM adoption.

---

[1]Graduate Student, Dept. of Civil and Environmental Engineering, University of Washington, Seattle, WA, USA, sangermd@uw.edu, Corresponding author
[2]Associate Professor, Dept. of Civil and Environmental Engineering, University of Washington, Seattle, WA, USA, bwmaurer@uw.edu

## Introduction

Historical trends and technological innovations can shape not only the content of research but also shift the way it is communicated in terms of vocabulary, tone, and style (e.g., Bochkarev et al., 2014; Zhou et al., 2023; Steuer et al., 2024). A striking example of this phenomenon is the emergence of large language models (LLMs), such as ChatGPT, that can generate and refine text with unprecedented fluency and scale. Since the launch of the first ChatGPT model in late 2022, LLMs have rapidly gained traction for performing a plethora of tasks relating to research formation, execution, and publication, including drafting and editing text, summarizing and critiquing documents, and even generating complete manuscripts (e.g., Májovský et al., 2023). Numerous beneficial uses of LLMs, as well as harmful misuses, have been documented and may be further imagined (e.g., Berdejo-Espinola & Amano, 2023; Lindsay, 2023; Van Noorden and Perkel, 2023; Mittelstadt et al., 2023; Lu et al., 2024). Given their transformative potential, a wealth of general and domain-specific analysis and commentary has rapidly emerged on LLMs, including research on their use in civil and environmental engineering (CEE) education, scholarship, and practice (e.g., Sanger and Maurer, 2023; Kumar, 2024; Plevris, 2025; Matzakos and Moundridou, 2025; Li et al., 2025).

Anecdotally, we observe that LLMs are used to some degree by virtually all CEE students to complete coursework, but also permeate the highest levels of research, where they are employed by journal authors and reviewers alike. However, it is unknown how prevalent LLM adoption is across CEE scholarship and at what rate authors disclose it, and detecting and quantifying the influence of LLMs in CEE scholarship is challenging. Many efforts to detect AI-generated text in other corpora have relied on methods that require very large curations of human- and LLM-authored texts (e.g., Guo et al., 2023; Li et al., 2024). Using such data, classification models based on AI architectures and traditional statistical methods have been trained (e.g., Wu, J., et al., 2025), but many of these models (i) lack interpretability (i.e., it is unclear what features drive classification decisions); (ii) are biased against non-native English writers (Liang et al., 2023); (iii) perform poorly outside the parameter space of their training data (e.g., Li et al., 2023); and (iv) can depend on the LLM model and version, as well as the prompts used to generate the LLM text. A detection model trained on text from GPT-2, for example, had greatly reduced efficacy on newer GPT versions (Bao et al., 2023). Given these constraints, attempts to quantify the prevalence of LLM usage across entire journals or research fields appear rare (e.g., Liang et al., 2024; Kobak et al., 2025). In a recent study, Kobak et al. (2025) applied a novel method inspired by the concept of excess mortality to a large corpus of biomedical research abstracts. By identifying significant shifts in the frequency of certain words starting in 2024, which they refer to as "excess vocabulary," Kobak et al. (2025) suggested lower-bound estimates on LLM usage that range from 5% to 40% across different countries and journals but argue "the true LLM usage in biomedical publishing may be closer to the highest lower-bounds we observed."

Inspired by the seminal work of Kobak et al. (2025), we study 25 years of publicly available abstracts from the portfolio of articles published in American Society of Civil Engineers (ASCE) journals and conference proceedings between 2000 and 2025. By studying linguistic trends over 25 years, we estimate the prevalence of LLM-generated writing and assess how its adoption varies across disciplines and publication types within the field of CEE scholarship, for which ASCE is chosen as a representative sample. While the results may not generalize perfectly to all publishers, the dataset offers a valuable case study for understanding LLM use within a broad sample of CEE scholarship. In addition to estimating the frequency of LLM use, we also investigate how LLM adoption is associated with changes in the structure (i.e., length and punctuation), complexity (i.e., word choice diversity and readability), voice (i.e., active or passive), and embedded confidence (i.e., use of qualifying or hedging language) of research prose. Importantly, the objective of this study is not to prescribe policy regarding LLMs in scholarly writing. Rather, the contribution is empirical: to provide the first large-scale, data-driven portrait of how the language of CEE research is changing alongside LLM adoption. Quantifying these trends establishes an evidence base that can inform discussions of authorship norms, disclosure requirements, and editorial policy.

## Data and Methodology

### *Corpus Development*

To prepare the textual dataset for this study, we compile the publicly available abstracts from January 2000 through December 2025 published by ASCE across 32 journals and 935 conference proceedings. The resulting corpus comprises 149,452 abstracts, distributed in terms of year and "venue" (i.e., journal or conference proceeding) as shown in Figure 1. In addition to the raw-text corpus, we construct a processed textual dataset to enable vocabulary- and frequency-based statistical linguistic analyses. Preprocessing is performed using standard natural language processing techniques implemented in Python. Specifically, each abstract is tokenized into individual word tokens using the Natural Language Toolkit (NLTK) (Bird et al., 2009), with all text converted to lowercase for case-insensitive analysis. Common stop words (e.g., "the," "and," "of"), which carry limited semantic content and can distort frequency-based measures, are removed using NLTK's curated stop-word lists. The remaining tokens are then lemmatized using NLTK's WordNet lemmatizer, informed by part-of-speech tags assigned in full-sentence context prior to stop-word removal, to reduce inflected word forms to their base or dictionary form; for example, "models," "modeling," and "modeled" are all mapped to the lemma "model." Derived words are not merged with their roots; for example, "enhancement" remains distinct from "enhance." As a simple illustration, the raw text "The dogs were running in the park, and they chased a cat to corner it near a tree" is reduced to "dog run park chase cat corner tree." The processed corpus is used for all analyses involving word frequencies, vocabulary shifts, and stylistic marker identification, while the raw-text corpus is retained for analyses requiring original sentence structure and word order.

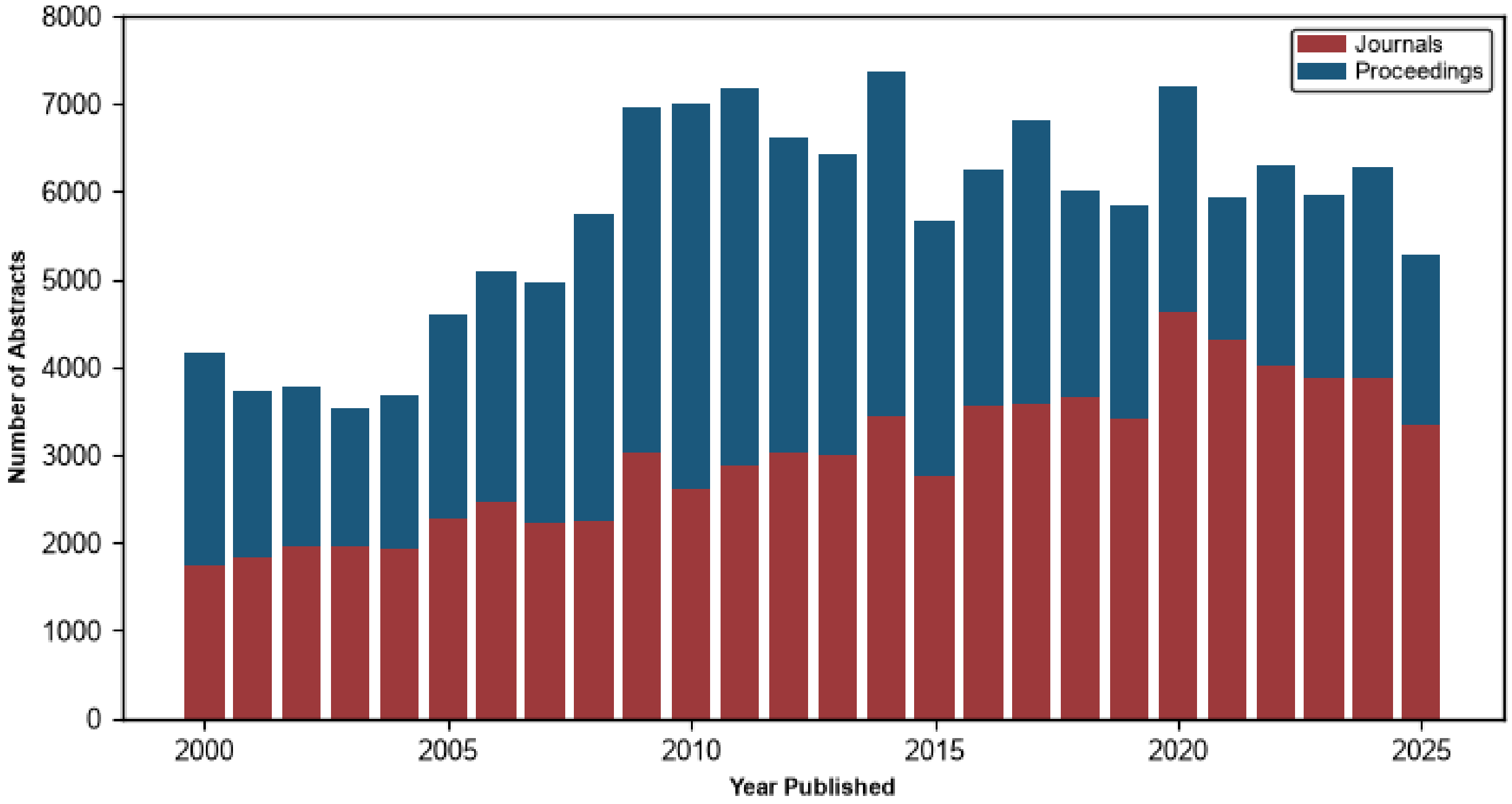

Figure 1. Distribution of abstracts analyzed.

***Semantic Properties***

The curated corpus provides a novel opportunity to investigate linguistic trends in CEE scholarship over the last 25 years and to assess whether LLM adoption coincides with departures from these trends. We quantify several semantic properties for each abstract, which together describe the structure, complexity, voice, and embedded confidence of prose. Overall abstract length in terms of words and sentences, mean sentence length, and punctuation use specify the structure of each abstract. Punctuation frequency, in addition to the other count-based metrics presented momentarily, is normalized by the number of sentences in the abstract to account for shifting length trends over time. Abstract complexity is enumerated in terms of word choice diversity and two commonly used readability scores (e.g., Wang et al., 2013): the Gunning Fog Index (Gunning, 1968) and Flesch-Kincaid Reading Level (Kincaid et al., 1975). Word choice diversity is defined as the number of unique words divided by the number of total words in an abstract. The Gunning Fog Index estimates the years of formal education required to understand text on first reading (Eq. 1):

$$Gunning\ Fog\ Index = 0.4 * \left(\left(\frac{Total\ words}{Total\ sentences}\right) + 100 * \left(\frac{Complex\ words}{Total\ words}\right)\right) \text{ (Eq. 1)}$$

where complex words are those containing three or more syllables, excluding proper nouns and common suffixes (e.g., -ing, -ed). The Flesch-Kincaid Reading Level similarly estimates the minimum reading level needed to comprehend the text on first reading (Eq. 2):

$$Flesch - Kincaid\ Reading\ Level = 0.39 * (M_W) + 11.8 * (M_S) - 15.59 \text{ (Eq. 2)}$$

where $M_W$ is the mean number of words in a sentence and $M_S$ is the mean number of syllables in a word. The Gunning Fog Index and Flesch-Kincaid Reading Level have their own suite of limitations, including but not limited to: (i) the limited empirical samples on which they were fit; (ii) their oversimplification of complexity in terms of syllable quantity and sentence length; and (iii) their lack of consideration for context and reader familiarity with the subject matter. However, rather than classifying a single text excerpt for a given reading level, their primary applications in this study are to observe relative complexity changes over time and to examine how LLM influence may affect complexity.

Anecdotally, we understand that norms around voice and verbiage have shifted through time in CEE scholarship, particularly as they relate to the use of active, first-person narrative (e.g., “We analyzed the data”) and passive narration (e.g., “The data were analyzed”). As proxy metrics to observe verbiage trends, we equate active voice to first-person pronoun use and passive voice to instances of “to be” verbs followed by past participles (e.g., “was determined,” “is being studied”). Similarly, we represent the relative confidence of research prose by tracking instances of cautious or “hedging” words (i.e., might, may, could, would, suggest, potentially, possibly, tend to, perhaps, likely, seem, and appear). As with the readability metrics, the instance tracking of first-person pronouns, past participles, and hedging words are likely incomplete assessments of active voice, passive voice, and confidence, respectively, in terms of absolute usage. However, they satisfy the objective of detecting relative changes through time for the purposes of this study.

***Linguistic Signals of LLM Usage***

To estimate the prevalence of LLM use, we adopt a frequency-shift methodology inspired by Kobak et al. (2025) that identifies statistically anomalous vocabulary changes. Using the processed textual dataset, we compute the annual document-level frequency of each lemma (i.e., word) as the proportion of abstracts each year containing at least one instance of the word normalized by the number of publications that year. This approach prevents longer abstracts and years with more abstracts from disproportionately affecting frequency estimates and ultimately yields a year-word frequency matrix spanning the temporal and lexical domain of the corpus. With this matrix, we observe trends in word usage through time and investigate deviations from their expected trends.

The launch of OpenAI’s GPT-3.5 model in November 2022 is largely considered the introduction of LLMs for widespread public use. In this frequency-shift methodology, we isolate all abstracts with a publication year between 2000 and 2022 as the pre-LLM dataset, and those with a publication year of 2024 or 2025 as the post-LLM dataset. The expected frequencies for each word are quantified using a counterfactual trendline, where a least-squares linear model is fit to the observed frequencies between 2010 and 2022 and extrapolated for comparison to the observed frequencies in 2024 and 2025. Deviations from the expected frequencies are computed according to two complementary departure metrics: frequency gap, $\delta$ (Eq. 3), and frequency ratio, $r$ (Eq. 4):

$$\delta(w, y) = f_y(w) - \hat{f}_y(w) \text{ (Eq. 3)}$$

$$r(w, y) = \frac{f_y(w)}{\hat{f}_y(w)} \text{ (Eq. 4)}$$

where $f_y(w)$ is the observed frequency and $\hat{f}_y(w)$ is the expected frequency based on the counterfactual trendline of word $w$ in year $y$. These two metrics are considered in tandem, as δ highlights the absolute change in prevalence and $r$ emphasizes the relative change in prevalence. Therefore, deviations identified by a large $\delta$ indicate changes in the use of common words with high baseline frequency, whereas departures identified by a large $r$ indicate changes in the use of rare words with low baseline frequency. Considering the top departures in 2025, for example, the largest deviations in terms of $\delta$ are "enhance," "offer," and "demonstrate," and the largest style-word deviations in terms of $r$ are "boast," "stride," and "underscore." Figure 2 presents the 15 largest $\delta$ deviations observed in 2025, annotated with the corresponding $r$. Many words exhibit dramatic spikes in use-frequency, which we interpret – on the whole – as evidence of LLM use. It is interesting to note that many of these words had been increasing in use, particularly in the recent decade, but their use increased sharply in 2024 and 2025, suggesting LLMs may amplify preexisting trends in human writing. While this study focuses on post-LLM departures from expected trends, we compute frequency departures for all publication years from 2005 to 2025, considering expectations based on the counterfactual trendline. The top 32 $\delta$ departures since 2005 occur in 2024 and 2025, such that the vocabulary changes coinciding with the emergence of LLMs are otherwise unprecedented in the recent history of CEE scholarship. We acknowledge that linguistic evolution may not be strictly linear, as implied by our linear modeling of counterfactual trajectories, but this approach provides a simple and transparent baseline against which abrupt departures can be identified. Alternative nonlinear models could be presented in future work but would not change the central observation that the magnitude and simultaneity of post-2023 vocabulary shifts are historically unprecedented within the corpus.

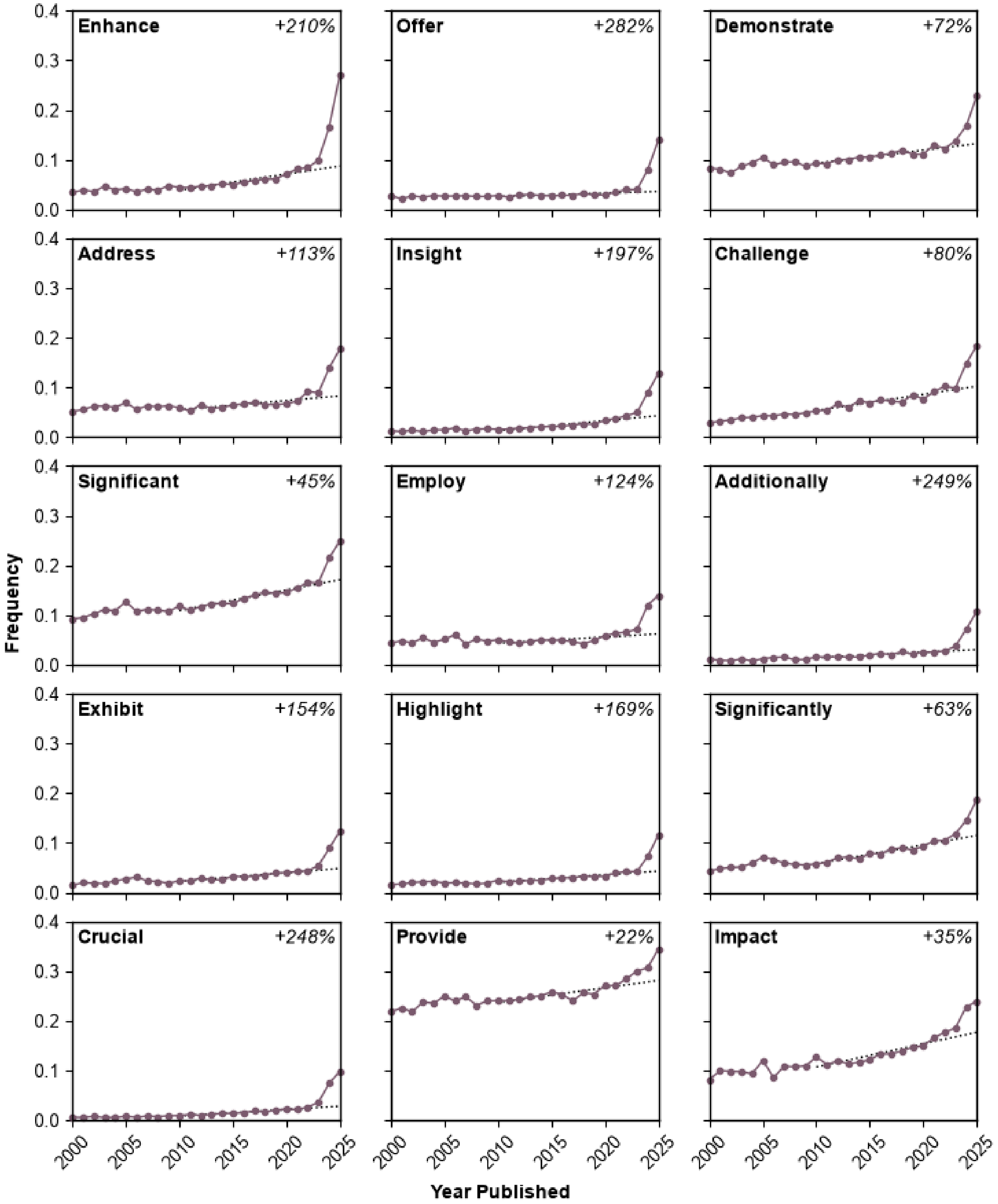

**Figure 2.** Largest frequency gap, $\delta$, departures in 2025, annotated with the percent deviation from the counterfactual trend.

Like Kobak et al. (2025), we identify preliminary thresholds for $\delta$ and $r$ for which words exceeding either threshold in 2025 are identified as potential "excess" words (i.e., words that may be indicative of LLM use) (Figure 3). The preliminary thresholds, $\delta = 0.03$ and $r$ as a function of 2025 frequency ($f$), $r(f) = f^{\log_{10} 3/-2}$ (Figure 3), are – in essence – tuned during a subsequent methodological step described momentarily. The initial list comprises 188 words, which are classified as either "style" words or "content" words. For example, "enhance" is classified as a style word, whereas "environmental" is classified as a content word. Table S1 in the *Supplemental Materials* presents the complete style and content classification of excess words. Only excess style words, of which there are 122, are considered in the LLM-use analysis (all words in Figure 2 are style words). The exclusion of content words intends to filter out vocabulary relating to research subject-matter and tools (e.g., "infrastructure," "neural," "climate"), that naturally fluctuate in use due to emergent interests, problems, and events. The distinction between style and content words necessarily involves some degree of judgment; the classifications given in Table S1 enable transparency and reproducibility but could be refined in future work.

Notably, the sudden observed increases in certain style words (Figure 2, Table S1), which are hypothesized to result from LLM use, greatly exceed the largest observed increases in content words, many of which are associated with some of the biggest emergent trends of the last 25 years. These include climate change ("sustainable," "sequestration," "permafrost," "arctic," "coral"), computing and sensing ("broadband," "digitalize"), machine learning ("model," "data," "transformer," "explainable," "decoder"), and new forms of AI ("generative"). "Sustainable" and "underscore," for example, have very similar background frequencies (i.e., $\hat{f}_y$ values); for "sustainable," the largest one-year deviation from this background was 96% which, while dramatic, may be unsurprising to those who performed research in the 2000s. Yet for "underscore," the 2025 deviation from this background was approximately 2600% ("underscore" is also among the top 70 words in terms of $\delta$ deviation). In isolation, such large anomalies might be reasoned to result from noise (i.e., unrecognized causes) or, in the case of extremely rare words, to be expected. But in totality, the many unprecedented, synchronous shifts in vocabulary share a plausible common cause – LLMs – the influence of which appears to greatly exceed the largest CEE research movements, tools, and events of the 21st century.

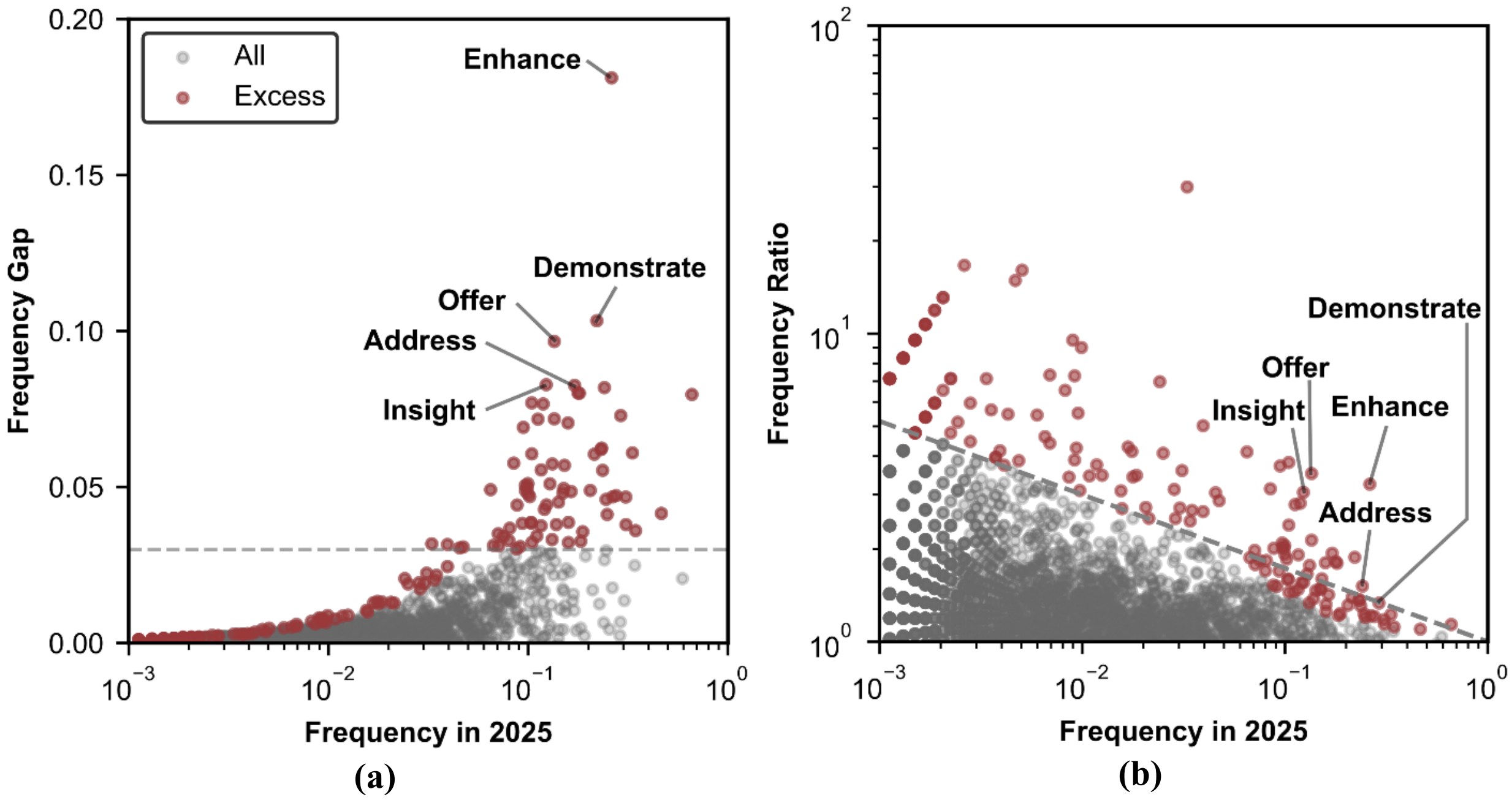

**Figure 3.** Initial identification of excess words identified in terms of frequency (a) gap, $\delta$, and (b) ratio, $r$, where the preliminary frequency thresholds are illustrated with the dotted lines and words classified as excess by either threshold are colored red in both panels.

The $\delta$ and $r$ thresholds are tuned independently to identify the subset of excess style words that best delineates abstracts published in post-LLM years (2024 and 2025) from those of pre-LLM publication (2022). For example, $r$ thresholds between 0.001 and 0.1 are used to iteratively define candidate subsets of rare excess style words, $G_{rare}$ (i.e., words with generally low background frequency that saw large relative changes in post-LLM use). For each candidate $r$ threshold and corresponding rare-word set $G_{rare}$, we compute the rare-word group-prevalence differential, $\Delta_{Grare}$ (Eq. 5):

$$\Delta_{Grare} = P_{Grare} - Q_{Grare} \text{ (Eq. 5)}$$

where $P_{Grare}$ is the fraction of post-LLM year abstracts containing at least one word in set $G_{rare}$, and $Q_{Grare}$ is the fraction of pre-LLM year abstracts containing at least one word in set $G_{rare}$. The optimal $r$ threshold is defined as the value that maximizes $\Delta_{Grare}$. Applying this procedure to 2024 abstracts, we find that an $r$ threshold of 0.04 best differentiates 2024 and 2022 abstracts, yielding a set of 66 rare words suggestive of LLM use. For this set, $\Delta_{Grare}$ between 2024 and 2022 is 17.2%, meaning 17.2% more abstracts in 2024 contain at least one of these rare-marker words than in 2022.

In parallel, we identify a subset of common excess style words, $G_{common}$ (i.e., words with generally high background frequency that saw large absolute changes in post-LLM use), via an analogous tuning procedure which maximizes the common-word group-prevalence differential $\Delta_{Gcommon}$. This exercise determines that a set of six common excess style words (enhance, employ, address, impact, utilize, and within) best differentiates 2024 and 2022 abstracts. Given this set of six common words, $\Delta_{Gcommon}$ between 2024 and 2022 is 13.3%, meaning 13.3% more abstracts in 2024 contain these common excess style words than in 2022. Like Kobak et al. (2025), we take the mean of the rare-marker and common-marker discrimination values (i.e., $\Delta_{Grare}$ and $\Delta_{Gcommon}$, respectively) as an estimate of LLM-influenced abstracts each year and thus estimate that 15.3% of abstracts in 2024 exhibit vocabulary shifts consistent with LLM usage.

This analysis is independently repeated to compare abstracts from 2025 to 2022. In 2025, a similar set of 69 rare words is identified using an optimum frequency-ratio threshold of 0.07, and, given this $G_{rare}$, 28.5% more abstracts contain these rare excess style words in 2025 than in 2022. In parallel, an optimal set of five common excess style words (enhance, offer, demonstrate, address, and insight) is found to best differentiate 2025 and 2022 abstracts, with 25.4% more abstracts in 2025 containing the common excess style words than in 2022. Therefore, our estimate of LLM-influenced abstracts in 2025 is 26.9%. The threshold-sweep figures from these analyses are provided in the *Supplemental Materials* for 2024 and 2025 as Figure S1 and Figure S2, respectively. Additional analysis and insights from these estimates are derived and discussed in the *Results and Discussion* section.

The preceding analyses apply the Kobak et al. (2025) framework as published to estimate the occurrence of LLM-influenced abstracts at the corpus level. We now extend that framework to examine how abstracts likely assisted by LLMs differ from historical linguistic trends beyond vocabulary. This extension requires a single metric to identify which abstracts in the CEE corpus are most likely influenced by LLMs. Towards this end, we construct a global set of LLM marker words by taking the union of the optimized rare- and common-marker sets identified for 2024 and 2025, comprising 69 rare and nine common marker words (Table S2). Because these words are used with some regularity before the introduction of LLMs, particularly the common-set words, classifying abstracts as LLM-influenced if they contain at least one word from this global set is not optimal. Ultimately, we classify an abstract as likely LLM-influenced if it contains at least two common stylistic markers and at least two rare stylistic markers from the global set (Table 1). These classification thresholds for rare and common marker counts are selected through sensitivity tests that evaluated false-positive rates prior to 2023 (i.e., the share of abstracts classified as LLM-influenced before LLMs existed, which serve as known negatives in the absence of ground-truth labels) against positive prediction rates in 2024 and 2025. We adopt the thresholds with the highest positive prediction rate for which the mean pre-2023 false-positive rate remains below 1%; these thresholds retain a classification rate of ~15%, which is similar to that estimated in the prior frequency-shift analyses, providing a measure of confidence in each approach (complete results from the sensitivity analyses are in Table S3). The results of this classification procedure are intended to capture population-level stylistic changes, and no claim is made towards LLM authorship of individual abstracts, which we opt not to identify. Abstracts classified as likely LLM-influenced are next used to describe the prevalence of LLM adoption and examine how the semantic properties of LLM abstracts compare against all others published over 25 years.

## Results and Discussion

### *Prevalence of LLM Adoption*

As just discussed, the frequency-shift methodology yields estimates of LLM adoption based on the mean prevalence differential of the rare and common word sets for the corresponding years: 15.3% in 2024 and 26.9% in 2025. Using the global marker word set derived from these two years, we apply a multi-marker classification procedure to isolate abstracts most likely assisted by LLMs. Applying this procedure produces another, more conservative estimate: 11.9% in 2024 and 18.8% in 2025 (Table 1). It must be emphasized in Table 1 that publications prior to 2023 comprise the baseline corpus representing the pre-LLM writing environment. These publications are included in Table 1 to elucidate false-positive prediction rates; the <1 % of pre-2023 abstracts classified as LLM-influenced are ignored in subsequent analyses. By identifying the abstracts most likely assisted by LLMs, we can describe which publication venues see the highest estimated rates of LLM-influenced abstracts. For example, there is a higher apparent rate in journals than in conference proceedings in 2024 and 2025 (Table 2). To quantify sampling uncertainty in these classification rates, we compute 95% confidence intervals using a two-stage cluster bootstrap with 5,000 replicates (Table 1, Table 2, and Table 3). In each replicate, publication venues (i.e., individual journals and conference proceedings) are first resampled with replacement, and abstracts are then resampled with replacement within each venue-year, so that abstracts from the same venue are not treated as independent observations. For the publication-type and discipline breakdowns (Table 2 and Table 3), venues are resampled within each group. The $2.5^{th}$ and $97.5^{th}$ percentiles of the resulting distribution define each interval. Although the point estimates for journals exceed those for conference proceedings in both 2024 and 2025, the corresponding 95% confidence intervals overlap in both years (Table 2), such that the apparent differences between the venues may be partly due to finite-sample uncertainty.

**Table 1.** Proportion of abstracts classified as LLM-influenced.

| Year | Number of Abstracts Classified as LLM-Influenced* | Percent of Total Abstracts | |
|---|---|---|---|
| | | Point Estimate | 95% Confidence Interval |
| 2000 | 9 | 0.2% | 0.0% - 0.5% |
| 2001 | 17 | 0.5% | 0.2% - 0.9% |
| 2002 | 8 | 0.2% | 0.0% - 0.5% |
| 2003 | 13 | 0.4% | 0.1% - 0.8% |
| 2004 | 10 | 0.3% | 0.1% - 0.6% |
| 2005 | 30 | 0.7% | 0.3% - 1.2% |
| 2006 | 25 | 0.5% | 0.2% - 0.9% |
| 2007 | 17 | 0.4% | 0.2% - 0.7% |
| 2008 | 30 | 0.6% | 0.3% - 0.9% |
| 2009 | 33 | 0.5% | 0.2% - 0.8% |
| 2010 | 39 | 0.6% | 0.3% - 1.0% |
| 2011 | 33 | 0.5% | 0.3% - 0.7% |
| 2012 | 38 | 0.6% | 0.3% - 0.9% |
| 2013 | 44 | 0.7% | 0.4% - 1.2% |
| 2014 | 51 | 0.7% | 0.4% - 1.1% |
| 2015 | 33 | 0.6% | 0.3% - 1.1% |
| 2016 | 49 | 0.8% | 0.4% - 1.2% |
| 2017 | 52 | 0.8% | 0.5% - 1.2% |
| 2018 | 51 | 0.9% | 0.5% - 1.4% |
| 2019 | 52 | 0.9% | 0.6% - 1.4% |
| 2020 | 77 | 1.1% | 0.7% - 1.6% |
| 2021 | 90 | 1.6% | 1.0% - 2.3% |
| 2022 | 94 | 1.5% | 1.0% - 2.2% |
| 2023 | 158 | 2.7% | 1.8% - 3.9% |
| 2024 | 731 | 11.9% | 9.6% - 14.9% |
| 2025 | 970 | 18.8% | 16.0% - 22.4% |

*__Note__: Years 2000 through 2022 are shown here for demonstration of false positive rate. The flagged abstracts from these years are ignored in the semantic trends analysis.

**Table 2.** Proportion of journal and conference proceedings abstracts classified as LLM-influenced.

| Year | Venue | Number of Abstracts Classified as LLM-Influenced | Percent of Total Abstracts | |
|---|---|---|---|---|
| | | | Point Estimate | 95% Confidence Interval |
| 2024 | Journal | 496 | 13.2% | 9.8% - 17.8% |
| | Proceedings | 235 | 9.9% | 6.8% - 13.1% |
| 2025 | Journal | 658 | 20.4% | 16.4% - 25.9% |
| | Proceedings | 312 | 16.0% | 12.4% - 20.1% |

We then categorize the 32 journals into a simplified subset of CEE disciplines according to the mapping schema of Table S4. Journals classified as pertaining to the construction domain exhibit the highest rate of LLM adoption, with an estimate of 34.6% in 2025, more than two and a half times the rate of journals classified to the geotechnical domain, which exhibit the lowest estimated rate of LLM adoption: 12.9% in 2025 (Table 3). In 2025, the 95% confidence interval for the geotechnical domain does not overlap that of the construction domain, indicating that this difference is statistically significant. The intervals of other discipline pairs in 2024 and 2025 do overlap, such that apparent differences between disciplines may partly reflect the small number of abstracts in some domains (e.g., coastal), and meaning that more data would be needed to establish significant differences between other disciplines, if present. Differences between domains, if persistent, may reflect several factors that the present data cannot disentangle, including differences in manuscript preparation norms, the proportion of industry-affiliated authors, the share of authors writing in English as an additional language, and review and publication timelines (e.g., longer timelines for geotechnical journals could produce an apparent lag in LLM-associated language).

The corpus-, venue-, and discipline-level estimates are striking but similar to those estimated in the biomedical research literature by Kobak et al. (2025). While alternative explanations could contribute to vocabulary changes over time (e.g., shifts in editorial practices, evolving disciplinary rhetoric), we hypothesize these would produce gradual evolution over time, and account for long-term changes via counterfactual trendlines. In contrast, the numerous vocabulary anomalies identified in this study appear abruptly. While these observations cannot prove that LLM use is the sole cause of the observed shifts, the timing of the deviations is consistent with the emergence of LLMs. The question then follows: if nearly one in five CEE abstracts shows signs of LLM influence, is this LLM influence changing the style of CEE scholarship? Towards this end, the 1,701 abstracts classified as likely LLM-influenced are isolated for comparative analysis.

**Table 3.** Proportion of domain-binned journal abstracts classified as LLM-influenced.

| Discipline | Year | Number of Abstracts Classified as LLM-Influenced | Percent of Journal Abstracts | |
|---|---|---|---|---|
| | | | Point Estimate | 95% Confidence Interval |
| Aerospace | 2024 | 12 | 9.2% | 4.6% - 14.5% |
| | 2025 | 23 | 22.1% | 14.4% - 29.8% |
| Architecture | 2024 | 11 | 14.5% | 6.6% - 22.4% |
| | 2025 | 13 | 27.1% | 14.6% - 39.6% |
| Coastal | 2024 | 4 | 12.9% | 3.2% - 25.8% |
| | 2025 | 5 | 14.3% | 2.9% - 25.7% |
| Construction | 2024 | 98 | 21.4% | 5.0% - 30.7% |
| | 2025 | 123 | 34.6% | 16.2% - 41.9% |
| Environmental | 2024 | 21 | 13.7% | 5.3% - 28.8% |
| | 2025 | 21 | 17.2% | 7.9% - 24.6% |
| General | 2024 | 185 | 14.4% | 9.6% - 33.6% |
| | 2025 | 259 | 22.5% | 16.6% - 33.6% |
| Geotechnical | 2024 | 54 | 9.1% | 6.7% - 12.8% |
| | 2025 | 61 | 12.9% | 8.2% - 16.4% |
| Hydrological | 2024 | 42 | 12.6% | 4.3% - 20.5% |
| | 2025 | 46 | 17.2% | 9.9% - 24.9% |
| Structural | 2024 | 43 | 8.8% | 4.8% - 15.8% |
| | 2025 | 65 | 13.2% | 8.6% - 19.3% |
| Transportation | 2024 | 26 | 12.7% | 3.2% - 18.5% |
| | 2025 | 42 | 24.6% | 13.6% - 32.2% |

***Semantic Trends and Influence of LLMs***

Using the defined semantic properties related to structure, complexity, voice, and embedded confidence of the research prose, we assess trends in CEE scholarship over the last 25 years and investigate how LLM writing may relate to these trends. We present and discuss the temporal trends of each semantic property as a mean line plot overlain by the mean property value of the abstracts classified as likely LLM-influenced in 2024 and 2025, which have been removed from the larger corpus. Results are bifurcated into journal and conference proceedings to investigate differences across venue. We also provide a complementary plot for each property in which the values are normalized to the mean value in the publication year 2000. This normalized representation highlights relative changes over time, illuminating patterns that can be challenging to see across the scale and range of the property value in absolute terms. Results for all properties across the entire corpus (including LLM-classified and unclassified abstracts) are provided in the *Supplemental Materials*; here we present and discuss select results.

To characterize uncertainty in each annual mean, we compute 95% confidence intervals via bootstrap sampling, shown as shaded bands for the background corpus and as error bars for the likely LLM-influenced abstracts (Figures 4-11). In many cases, these intervals are so narrow that they are difficult to distinguish from the plotted lines and markers (e.g., Figures 6-8). For each year, the abstracts are resampled with replacement 5,000 times, and the 2.5$^{th}$ and 97.5$^{th}$ percentiles of the resampled means define the interval. In the normalized plots, the year-2000 mean used for normalization is also resampled in each replicate, so that its uncertainty carries into the normalized values. Where the intervals of the likely LLM-influenced abstracts and the background corpus do not overlap, we interpret the difference as statistically significant. As will be shown, several semantic properties show apparent deviations from their recent trends beginning around 2023. When isolating the likely LLM-influenced abstracts from the broader corpus, however, we see how these deviations are – in large part or entirely – concentrated in LLM-influenced abstracts. In this way, presenting the mean value illustrates the apparent influence of LLMs on CEE scholarship.

Before presenting the semantic-feature trends, we first show trends in authorship as an example of the framing applied throughout this section. The mean number of authors has risen steadily over time, from an average of approximately two in 2000 to nearly four in 2025 (Figure 4, Figure S3). The consistent increase has several possible causes, including: (i) the ever-increasing reach, speed, and capacity of the internet to facilitate collaboration; (ii) growing content complexity and domain specialization within CEE, and a corresponding need for more collaboration; and (iii) large cross-disciplinary centers of CEE scholarship (e.g., DesignSafe, Rathje et al., 2017), and the dissemination of work from those centers. In recent years, conference proceedings have, on average, fewer authors than journal articles and have exhibited less change over time. This may result from some combination of short timeline constraints when preparing conference proceedings, differences in minimum contribution for authorship norms, or differences in scope (i.e., proceedings may present work with smaller scope, managed by fewer authors). As may be expected, the number of authors per publication has no apparent relationship with LLM use.

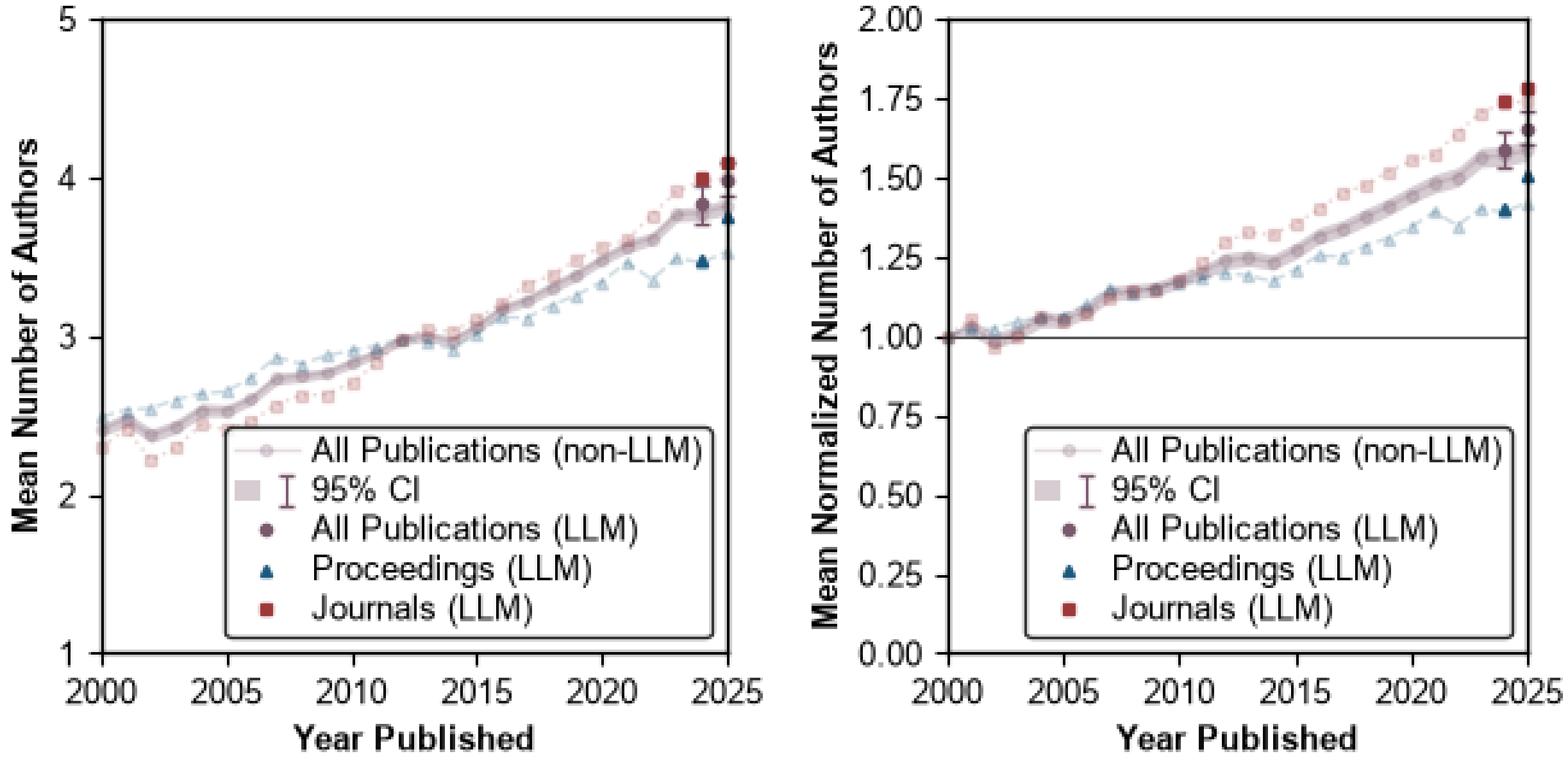

**(a)** **(b)**

**Figure 4.** The (a) mean number of authors per publication from 2000 to 2025 (b) normalized to the corresponding mean value in 2000, showing the LLM-influenced abstracts compared to CEE scholarship trends. Shaded bands and error bars denote 95% confidence intervals of the annual means for the background corpus and the likely LLM-influenced abstracts, respectively.

With this framing, we next consider abstract length, sentence length, and punctuation usage to describe how structural properties have evolved over time. Mean abstract length has risen, from 150 words and nearly seven sentences in 2000, to 200 words and over eight sentences in 2025 (Figure S4). Conference abstracts have, on average, shown less change over time, while journal abstracts have continued to lengthen steadily (Figure S4). Similarly, mean sentence length in journal abstracts has increased from approximately 23 words per sentence in 2000 to 25 words per sentence in 2025 (Figure 5, Figure S4). In contrast, LLM-influenced abstracts exhibit shorter sentences than the broader corpus, averaging approximately 24 words per sentence in 2024 and 2025, roughly one word shorter than other abstracts published in those years, with non-overlapping 95% confidence intervals indicating a statistically significant difference (Figure 5).

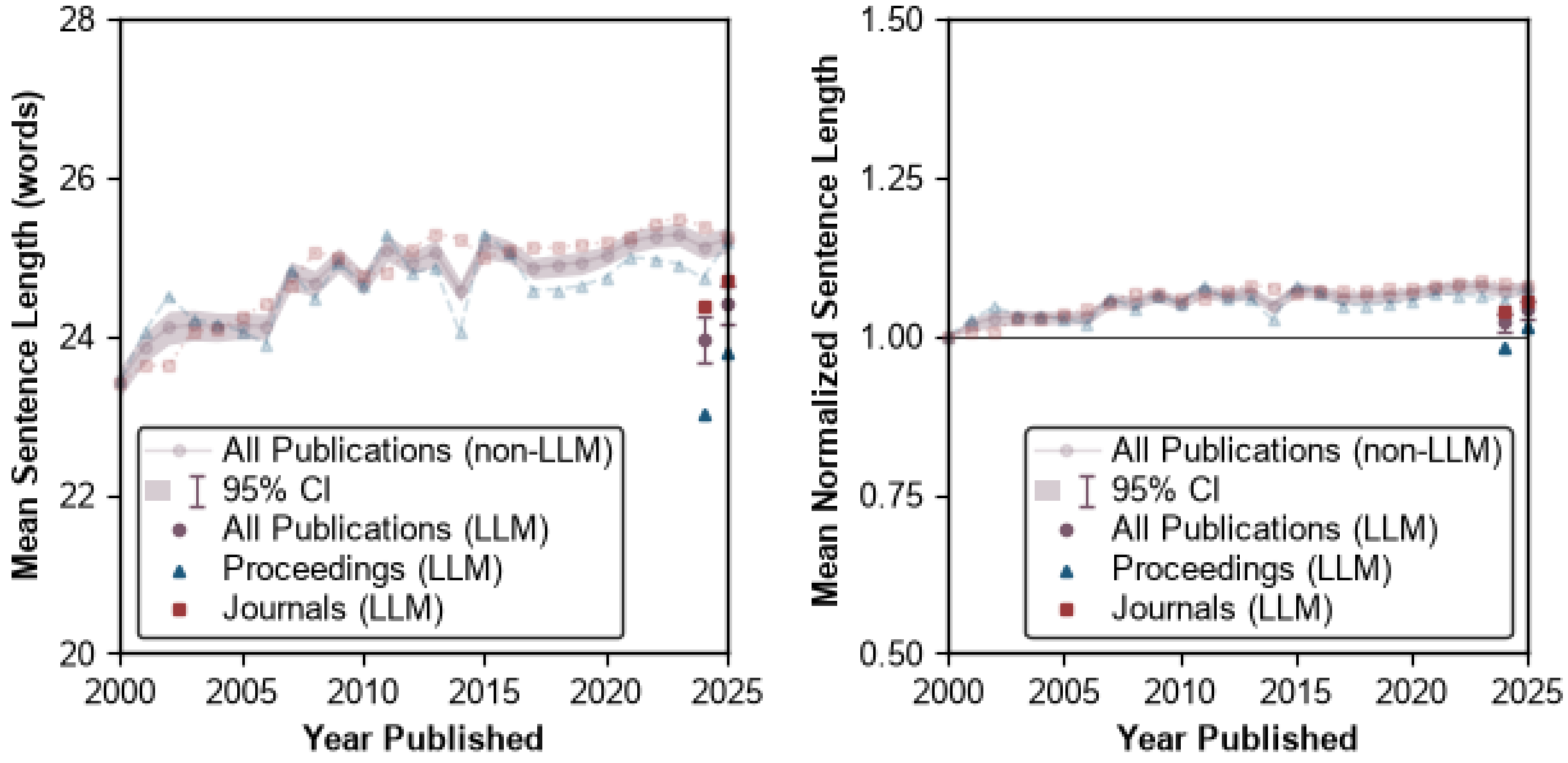

**(a)** **(b)**

**Figure 5.** The (a) mean sentence length in terms of number of words (b) normalized to the corresponding mean value in 2000, showing the LLM-influenced abstracts compared to CEE scholarship trends. Shaded bands and error bars denote 95% confidence intervals of the annual means for the background corpus and the likely LLM-influenced abstracts, respectively.

Punctuation provides an additional metric of how information is organized and communicated within sentences. Across the corpus, the use of em dashes, percent signs, and (to a lesser degree) commas has increased gradually over time, with a more pronounced rise beginning around 2015 (Figure 6, Figure S8). This trend is consistent with longer sentences and increased complexity, as punctuation such as commas and em dashes are used to organize ideas and clarify meaning. We anticipated increased em dash use in LLM-influenced abstracts, due to its broad association with LLM-generated text (e.g., Abebe, 2025; Csutoras, 2025; Wu, 2025); however, no clear deviation from the background trend is observed, potentially reflecting differences in the textual encoding of em dashes across source formats. The GPT series of LLMs, for example, is known to produce a style of em dash unbounded by spaces (e.g., "text—text") whereas most human typists opt for a shorter, space-bounded em dash (e.g., "text – text") (Abebe, 2025), but when these GPT-specific em dashes appear in publications, they may be altered by the publisher's textual encoding and thus may be impossible to distinguish from others. LLM-influenced abstracts do, however, exhibit significantly higher comma use relative to the broader corpus, suggesting a tendency toward more heavily segmented sentence structures (Figure 6). Such structures can contribute to complexity or improve readability, depending on the case; therefore, we seek to quantify these attributes more directly.

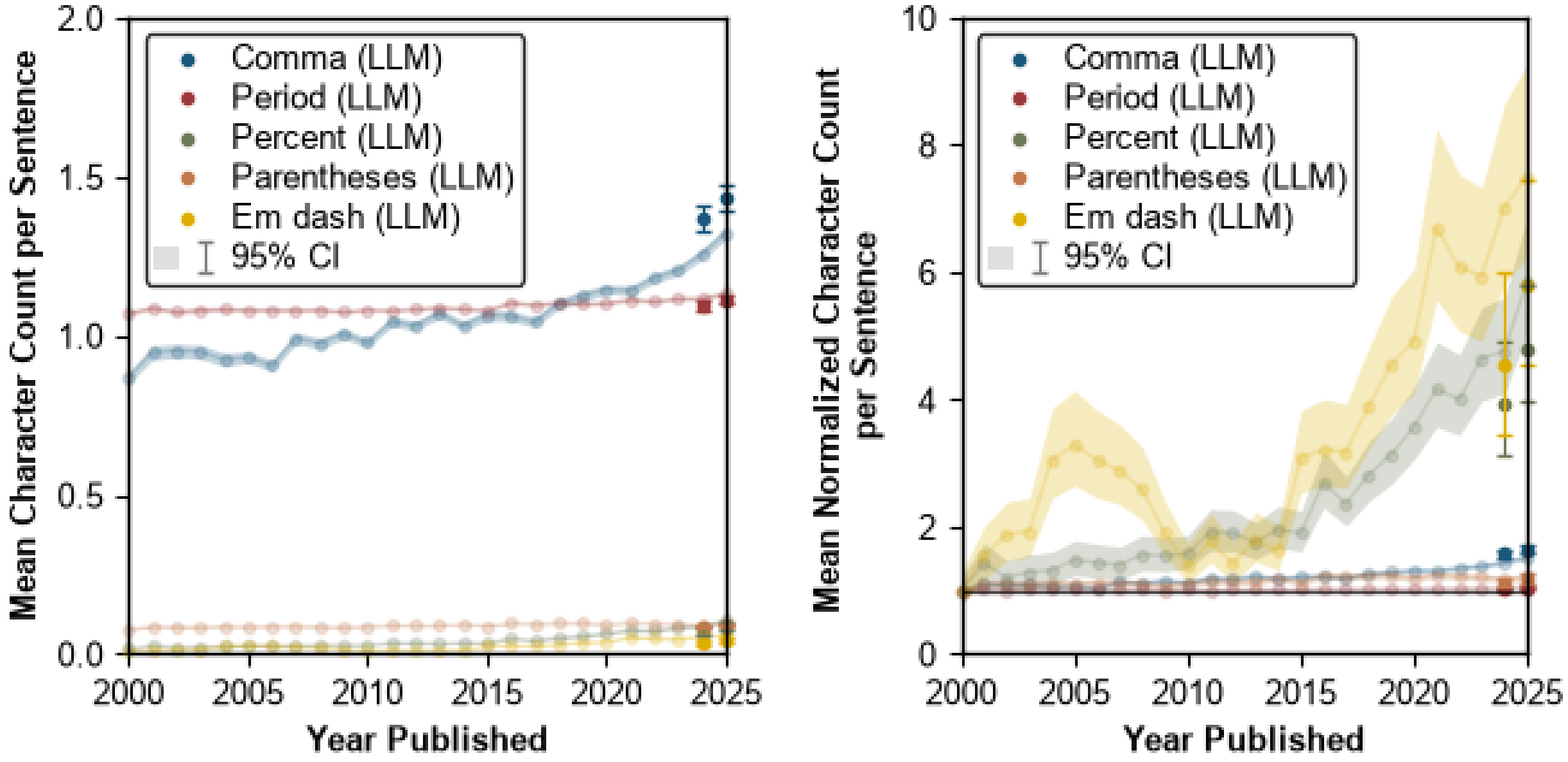

**(a)** **(b)**

**Figure 6.** The (a) mean character count per sentence of the five most frequently used punctuation marks in 2025 (b) normalized to the corresponding mean value in 2000, showing the LLM-influenced abstracts compared to CEE scholarship trends. Shaded bands and error bars denote 95% confidence intervals of the annual means for the background corpus and the likely LLM-influenced abstracts, respectively.

Abstract complexity in terms of mean word choice diversity declined gradually over most of the study period, from approximately 0.62 in 2000 to 0.57 in 2023 (Figure 7, Figure S5). Historically, conference abstracts have exhibited slightly greater lexical diversity than journal abstracts (possibly indicative of less editing), a difference of roughly 0.03 that has persisted throughout the study period. In 2024 and 2025, this long-term decline sharply reverses (Figure S5), and – when LLM-influenced abstracts are isolated – the observed increase appears associated with likely LLM-generated text to a statistically significant degree (Figure 7). Notably, word choice diversity among the background abstracts also departs from its prior downward trajectory in 2024 and 2025. This residual uptick may reflect (i) LLM-assisted abstracts that the intentionally conservative multi-marker classification procedure does not capture (e.g., those revised with LLMs but lacking the global marker words) and therefore remain among the unclassified abstracts, (ii) indirect LLM influence on human writing, or (iii) other factors that this analysis cannot distinguish. This suggests that LLM use may be reversing the gradual, decades-long decline in lexical diversity of CEE abstracts toward more varied word choice. Readability metrics further contextualize changes in abstract complexity by estimating the level of formal education required to understand the text on first reading. Both the Gunning Fog Index and Flesch-Kincaid Reading Level indicate that, on average, CEE abstracts in 2000 required a minimum of early college-level reading proficiency. Since 2000, the required reading grade level has increased (i.e., readability has declined) steadily over time toward a minimum reading level associated with college- to graduate-level education (Figure 8, Figure S6, Figure S7). LLM-influenced abstracts exhibit significantly higher required reading grades than the broader corpus, indicating increased complexity and reduced readability (Figure 8). As with other structural and semantic properties, this suggests that recent deviations in readability trends are disproportionately concentrated in likely LLM-influenced abstracts, and that LLM use may be associated with more complex and potentially less accessible research prose in CEE scholarship.

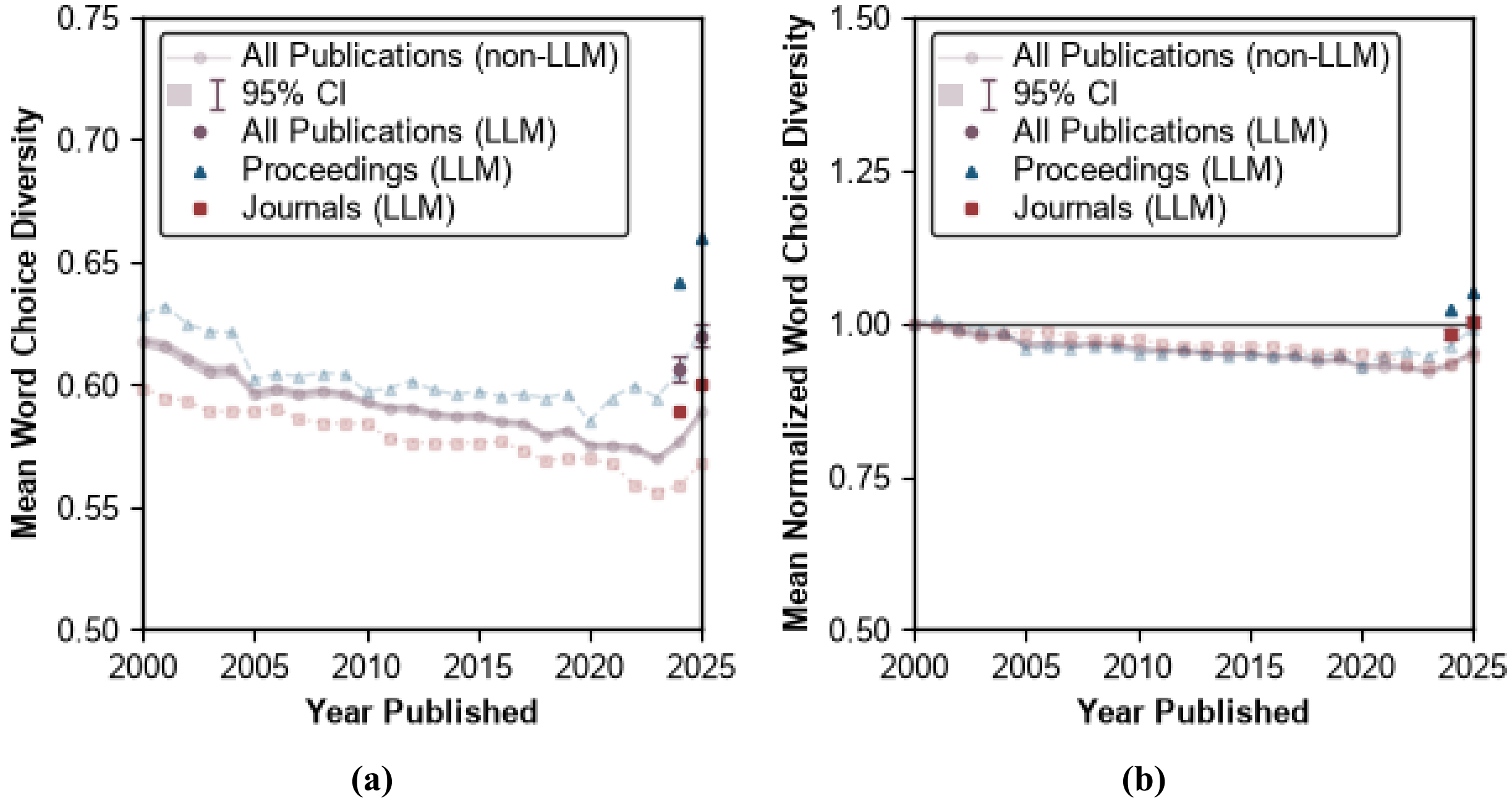

**Figure 7.** The (a) mean word choice diversity per abstract (b) normalized to the corresponding mean value in 2000, showing the LLM-influenced abstracts compared to CEE scholarship trends. Shaded bands and error bars denote 95% confidence intervals of the annual means for the background corpus and the likely LLM-influenced abstracts, respectively.

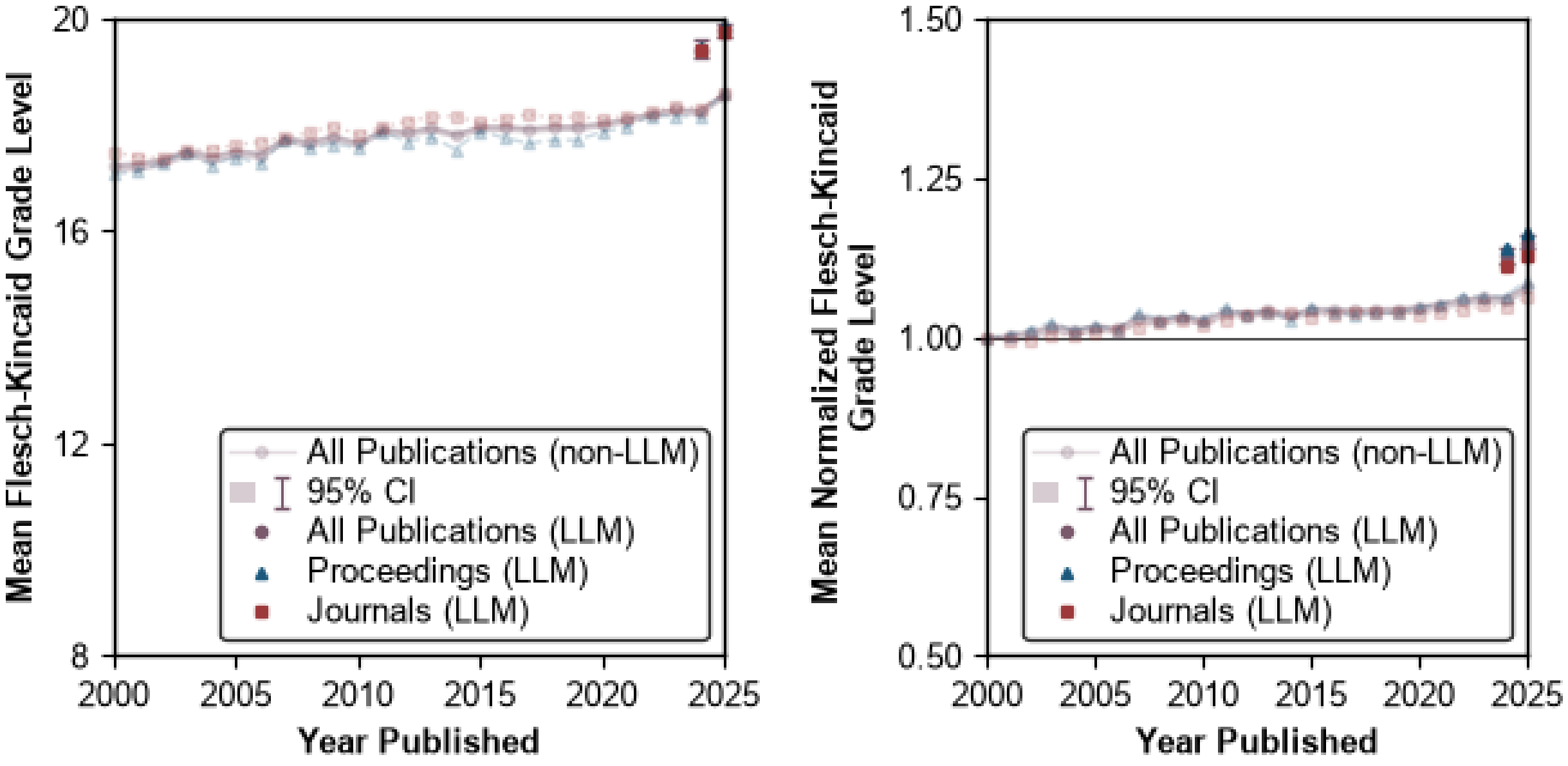

**(a)** **(b)**

**Figure 8.** The (a) mean Flesch-Kincaid Reading Level per abstract (b) normalized to the corresponding mean value in 2000, showing the LLM-influenced abstracts compared to CEE scholarship trends. Shaded bands and error bars denote 95% confidence intervals of the annual means for the background corpus and the likely LLM-influenced abstracts, respectively.

Beyond structural features, we examine trends in active voice, passive voice, and qualifying language as these choices, related to grammatical voice and rhetorical confidence, shape the perspective and certainty of research prose. Active voice, proxied via first-person pronoun use, and passive voice, proxied via past participles, have remained relatively constant over the last 25 years (Figure 9, Figure 10, Figure S9, Figure S10). The primary exception occurs in journal abstracts over the last decade, where an increase in active, first-person pronoun use is observed, likely reflecting changing norms and acceptance of first-person pronouns (e.g., "we") in CEE writing (Figure 9, Figure S9). Likely LLM-influenced abstracts also exhibit significantly higher first-person pronoun use than the broader corpus, particularly in 2024 (Figure 9). Congruently, beginning in 2024, a clear drop in the frequency of passive voice is observed (Figure S10), but the change in verbiage seemingly does not drive a 1:1 corresponding increase in first-person pronoun use (Figure S9). This might mean, as an example, that phrases such as "the model performance was evaluated" or "we evaluated the model performance," are written more directly, such as "the model performs in X manner" or "the model performance indicates X". When likely LLM-influenced abstracts are isolated from the broader corpus, the observed reduction in passive voice appears associated with the isolated abstracts to a statistically significant degree, suggesting that LLM use may be associated with this stylistic transition (Figure 10). Notably, qualifying, or "hedging," language also exhibits a pronounced decline beginning around 2024 (Figure 11, Figure S11), meaning results are less frequently qualified or caveated (e.g., "could," "might," "possibly"). As with passive voice and word choice diversity, this shift is statistically significant and appears concentrated in likely LLM-influenced abstracts (Figure 11), but a smaller residual decline is also evident among the background abstracts, which may similarly reflect LLM-assisted abstracts not captured by the classification procedure. Taken together, these findings suggest that LLM use is associated with more active grammatical constructions and with scholarly rhetoric that conveys greater author confidence and less uncertainty about research outcomes and significance. This could have important implications for how other researchers, policy makers, and the public understand, react to, and implement CEE research findings.

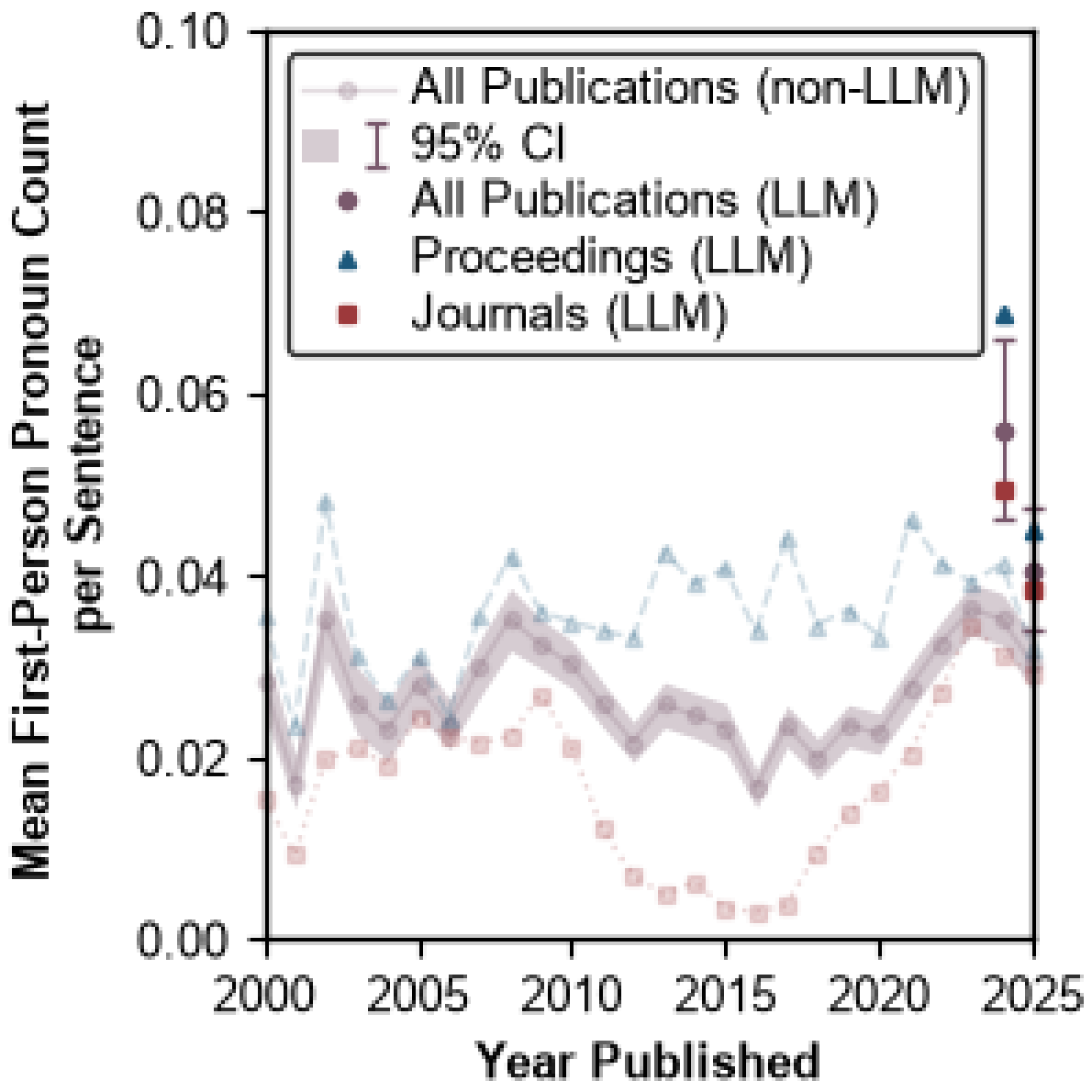

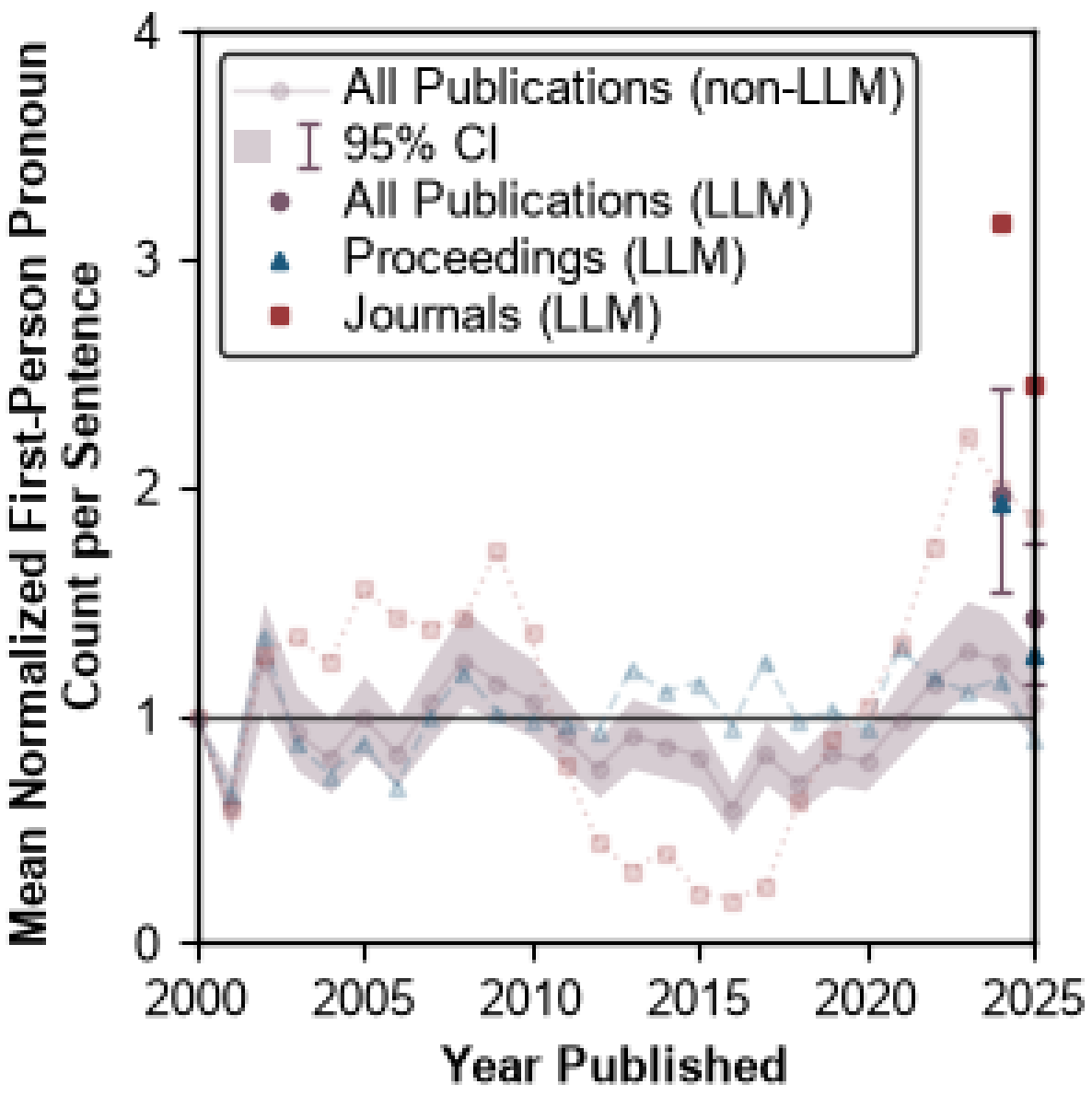

**(a)** **(b)**

**Figure 9.** The (a) mean first-person pronoun count per sentence (b) normalized to the corresponding mean value in 2000, showing the LLM-influenced abstracts compared to CEE scholarship trends. Shaded bands and error bars denote 95% confidence intervals of the annual means for the background corpus and the likely LLM-influenced abstracts, respectively.

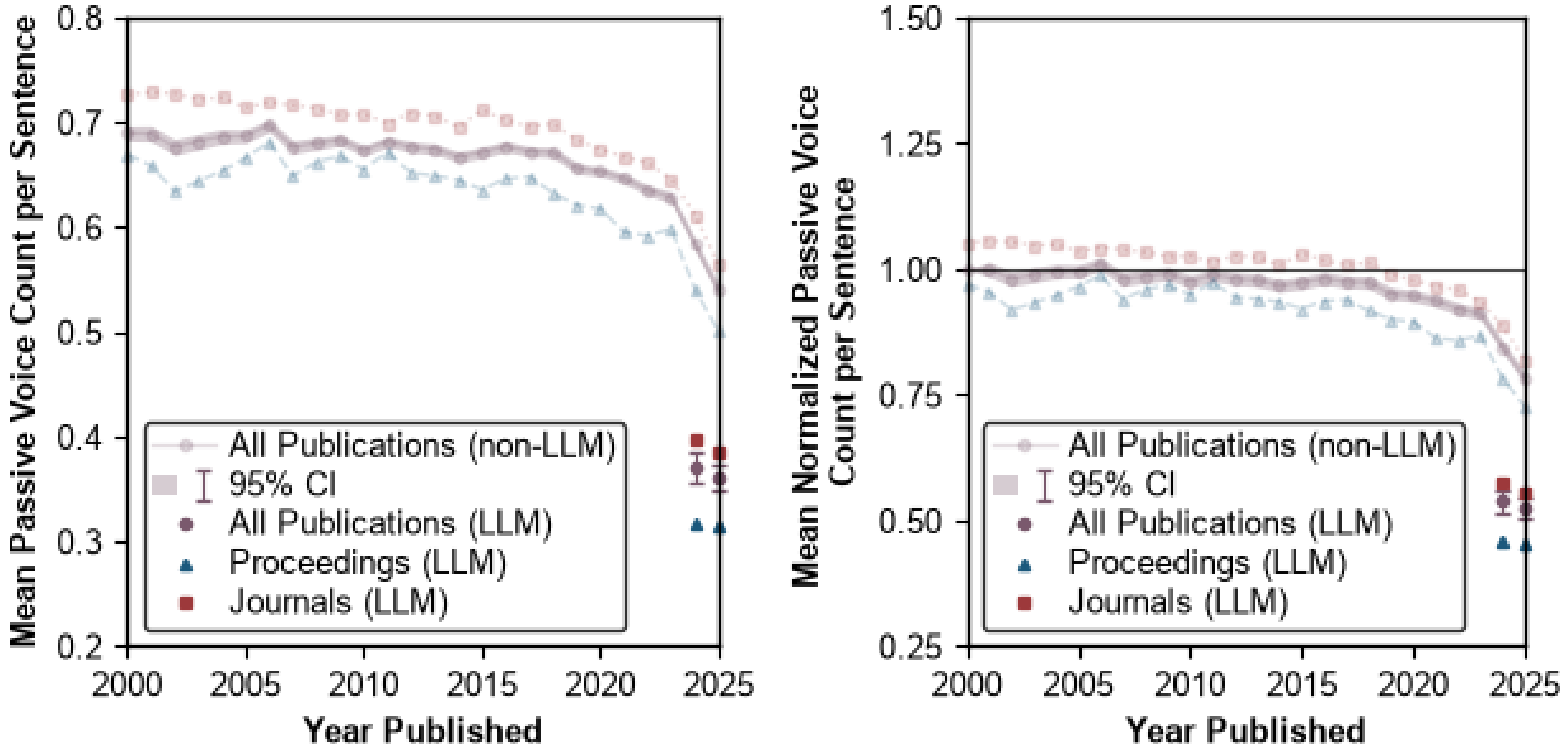

**(a)** **(b)**

**Figure 10.** The (a) mean passive voice count per sentence (b) normalized to the corresponding mean value in 2000, showing the LLM-influenced abstracts compared to CEE scholarship trends. Shaded bands and error bars denote 95% confidence intervals of the annual means for the background corpus and the likely LLM-influenced abstracts, respectively.

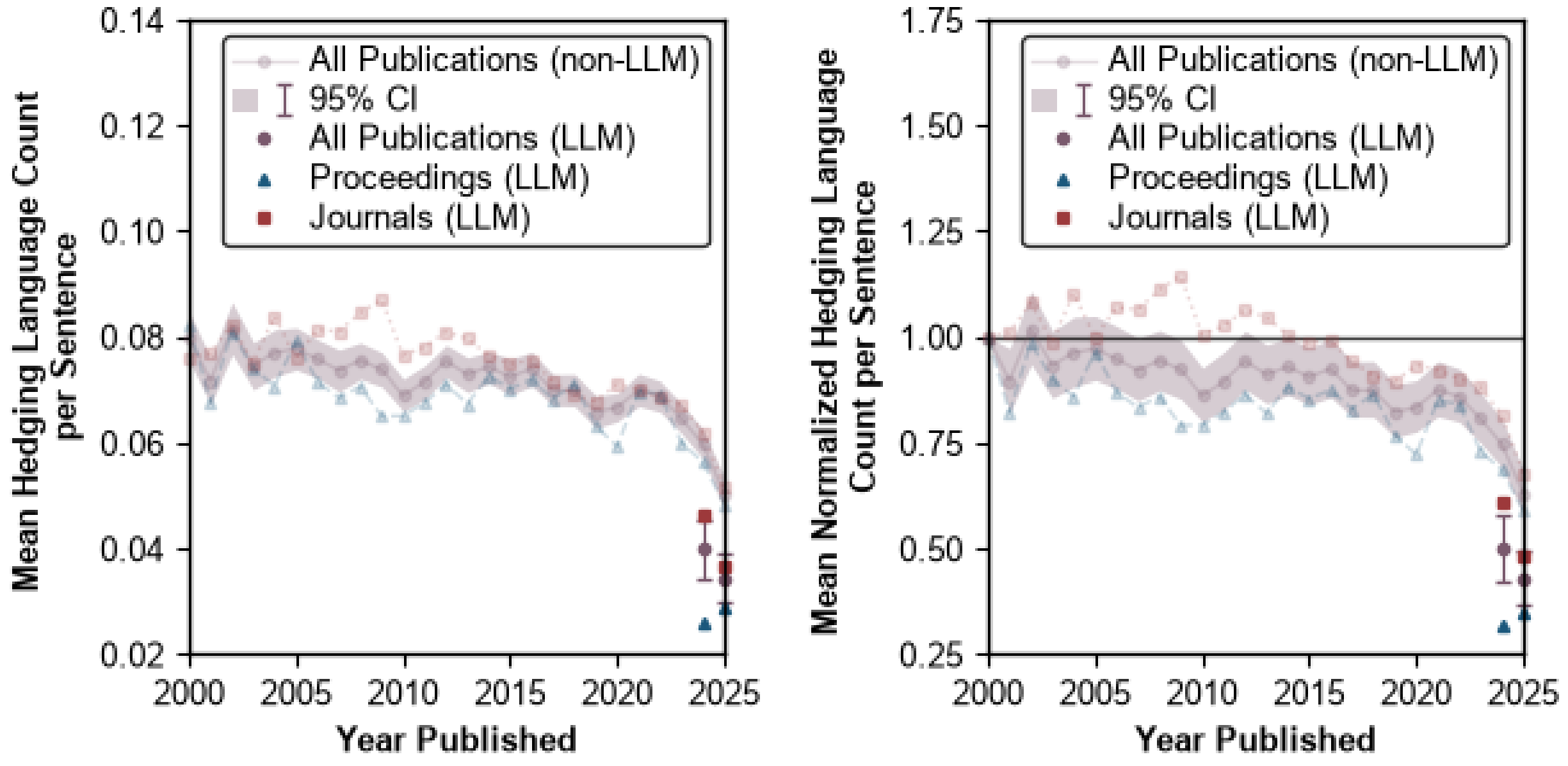

**(a)** **(b)**

**Figure 11.** The (a) mean hedging language count per sentence (b) normalized to the corresponding mean value in 2000, showing the LLM-influenced abstracts compared to CEE scholarship trends. Shaded bands and error bars denote 95% confidence intervals of the annual means for the background corpus and the likely LLM-influenced abstracts, respectively.

### *Limitations*

The findings of this study should be interpreted in the context of several limitations, some of which are previously discussed and are summarized here. *First*, and most notably, no ground-truth dataset exists identifying which abstracts were written with or without LLM assistance. Our methodology thus focuses on detecting population-level linguistic signals; these signals are used to make inferences about LLM use that are necessarily uncertain. *Second*, abstracts are a particularly efficient entry point for LLM assistance because of their summarization and synthesis functions. Thus, the classification of an abstract as likely LLM-influenced does not imply LLM authorship of the manuscript, which is not examined here nor feasible, as the manuscripts are not publicly available. *Third*, the corpus is also not balanced across disciplines as each discipline is represented in proportion to its publication volume in the venues studied. Because baseline use of the marker words may differ between disciplines, and disciplines with fewer abstracts yield less precise estimates, discipline-level prevalence estimates and comparisons will benefit from more data and analyses. ASCE venues may differ from other publishers in author background, geographic distribution, editorial policy, and abstract requirements, and thus the discipline-level findings may not generalize beyond ASCE. *Fourth*, because the multi-marker classification is intentionally conservative, some LLM-assisted abstracts likely remain among the unclassified abstracts, as suggested by residual departures in several semantic properties. This limitation is compounded by the rapid evolution of LLMs themselves, where differences between years may reflect both adoption and model changes. The marker words identified in this study may therefore become less reliable when extrapolated to future years, as newer models with different writing styles become available. *Finally*, the annual distributions of semantic properties are summarized by their means and bootstrapped confidence intervals; other features of these distributions (e.g., variance or skewness) could be investigated in future work.

## Conclusions

This study compiles publicly available abstracts published across ASCE journals and conference proceedings from 2000 through 2025 and applies a frequency-shift methodology to quantify LLM prevalence. Using this approach, estimates of LLM-associated linguistic patterns are 15.3% in 2024 and 26.9% in 2025. A more conservative multi-marker classification procedure, based on a global marker word set derived from these years, yields corresponding estimates of 11.9% and 18.8% of abstracts likely influenced by LLMs. Using the latter approach, 2025 estimates within specific domains of CEE research are as high as 34.6% and as low as 12.9%. These estimates provide the quantitative context necessary for interpreting stylistic changes associated with LLMs.

Several clear conclusions emerge regarding how research prose is changing alongside LLM use. Prior to the introduction of LLMs, CEE scholarship exhibits long-term trends toward increasing numbers of authors, longer abstracts, greater use of segmenting punctuation, higher required reading levels, and a shift toward active, first-person verb constructions, although many of these trends stabilize in recent years. Beginning around 2023, however, many excess style words (e.g., enhance, offer, demonstrate) dramatically depart from their historical trajectories, often greatly exceeding deviations associated with major emergent research movements and challenges in the 21st century. Multiple semantic properties also depart from historical trends beginning in 2024. When abstracts identified as likely LLM-influenced are isolated, these departures are shown to be largely or entirely concentrated in these abstracts. Specifically, abstracts classified as likely LLM-influenced exhibit systematic shifts, including increased word choice diversity, heavier use of commas, reduced readability, decreased reliance on passive constructions, and diminished use of qualifying language used to convey uncertainty. These features collectively suggest prose that is more segmented, more syntactically complex, and more assertive, indicating that LLM adoption is coinciding with a measurable and unprecedented stylistic evolution of scholarly writing in civil and environmental engineering. Whether these changes are "enhancing" scholarly communication is left to the reader to decide.